\documentclass{article}
\PassOptionsToPackage{table}{xcolor}
\usepackage{iclr2027_conference,times}
\usepackage[utf8]{inputenc}
\usepackage[T1]{fontenc}
\usepackage{hyperref,url,booktabs,amsfonts,amssymb,amsmath,microtype,xcolor}
\usepackage{graphicx,caption,subcaption,array,colortbl,placeins}
\usepackage[nameinlink,capitalise,noabbrev]{cleveref}
\definecolor{lm_purple}{HTML}{F1ECFA}
\newcolumntype{H}{>{\columncolor{lm_purple}}c}
\hypersetup{colorlinks=true,linkcolor=blue,citecolor=blue,urlcolor=blue}
\title{scTREAT: Latent Diffusion for\\Single-Cell Drug-Response Prediction}
\newcommand{\supmark}[1]{\textsuperscript{\normalfont #1}}
\newcommand{\authname}[2]{\textbf{#1}\supmark{#2}}
\author{%
  \begin{tabular}[t]{@{}l@{}}
  \authname{Zhaokang Liang}{*,1},\ 
  \authname{Shuyang Zhuang}{*,1},\ 
  \authname{Xiaoran Jiao}{*,1},\ 
  \authname{Weian Mao}{*,2} \\
  \authname{Hao Chen}{\textdagger,1},\ 
  \authname{Chunhua Shen}{\textdagger,1} \\
  {\normalfont\mdseries
    \supmark{1}Zhejiang University, China ; \ \supmark{2}Massachusetts Institute of Technology, USA}
  \end{tabular}
}
\iclrfinalcopy
\begin{document}
\maketitle
\begingroup
\renewcommand{\thefootnote}{\fnsymbol{footnote}}
\footnotetext[1]{Equal contribution.}
\footnotetext[2]{Corresponding authors.}
\endgroup
\pagestyle{plain}

\begin{abstract}
Generating drug-induced transcriptional responses requires a model to account for cellular background together with molecular identity and dose. We present scTREAT, the first diffusion framework to factorize state and drug guidance for single-cell drug-response prediction. A shared latent expression space connects baseline and response, while Guided Decomposable Attention (GD-Attn) injects a baseline representation and a dose-conditioned molecular embedding. Independent channel dropout exposes four conditioning states during training; inference combines three conditional predictions with separately specified state and drug coefficients, and dose modulates drug-guidance strength. On condition-level pseudobulk profiles derived from the Tahoe-100M single-cell atlas, scTREAT achieves the best value on 10 of 12 metrics for unseen cell line--drug combinations (UC) and 11 of 12 for unseen drugs (UD). Ablations examine condition injection, molecular representation, and guidance design. On UD, the relative gains over the second-best model reach 36.11\% and 34.21\% for DEG logFC-Spearman and logFC-Pearson. These results demonstrate the effectiveness of combining shared latent expression modeling with explicit state and drug conditioning for response prediction.
\end{abstract}

\section{Introduction}
Predicting transcriptional responses to unmeasured drug treatments requires modeling how cellular background, molecular identity, and dose jointly shape expression. Single-cell perturbation screens \citep{Klein2015inDrop,Datlinger2017_CROPseq,Srivatsan2020_sciPlex} connect compounds to transcriptional readouts and support mechanism-of-action (MoA) analysis and compound evaluation \citep{Lamb2006ConnectivityMap,Subramanian2017L1000,Trapotsi2022MoAReview,Ye2018DRUGseq,Corsello2020PRISM}. Yet the cost of screening limits coverage of the cell line $\times$ drug $\times$ dose space \citep{Mayr2009_NovelTrendsHTS,Macarron2011HTSImpact,Iversen2012_HTSValidation_AGM}. Conditional response generation offers a way to model the effects of candidate molecules across these unmeasured settings.

The modeling question is how to combine cellular background with treatment information and retain access to their contributions during inference. Latent perturbation models, molecular conditioning, and Transformers offer complementary approaches \citep{Lotfollahi2019scGen,hetzel2022predicting,Qi2024PRnet,Adduri2025.06.26.661135}. Diffusion-based methods have also established conditional generation and latent manipulation for single-cell data \citep{Tang2023scDiff,Luo2024scDiffusion,Sadria2025scVAEDer,He2026Squidiff,Shang2025dbDiffusion}. Within this line of work, we study a shared conditional generator whose state and drug guidance can be weighted separately at inference.

\begin{figure*}[!t]
  \centering
  \includegraphics[width=\linewidth]{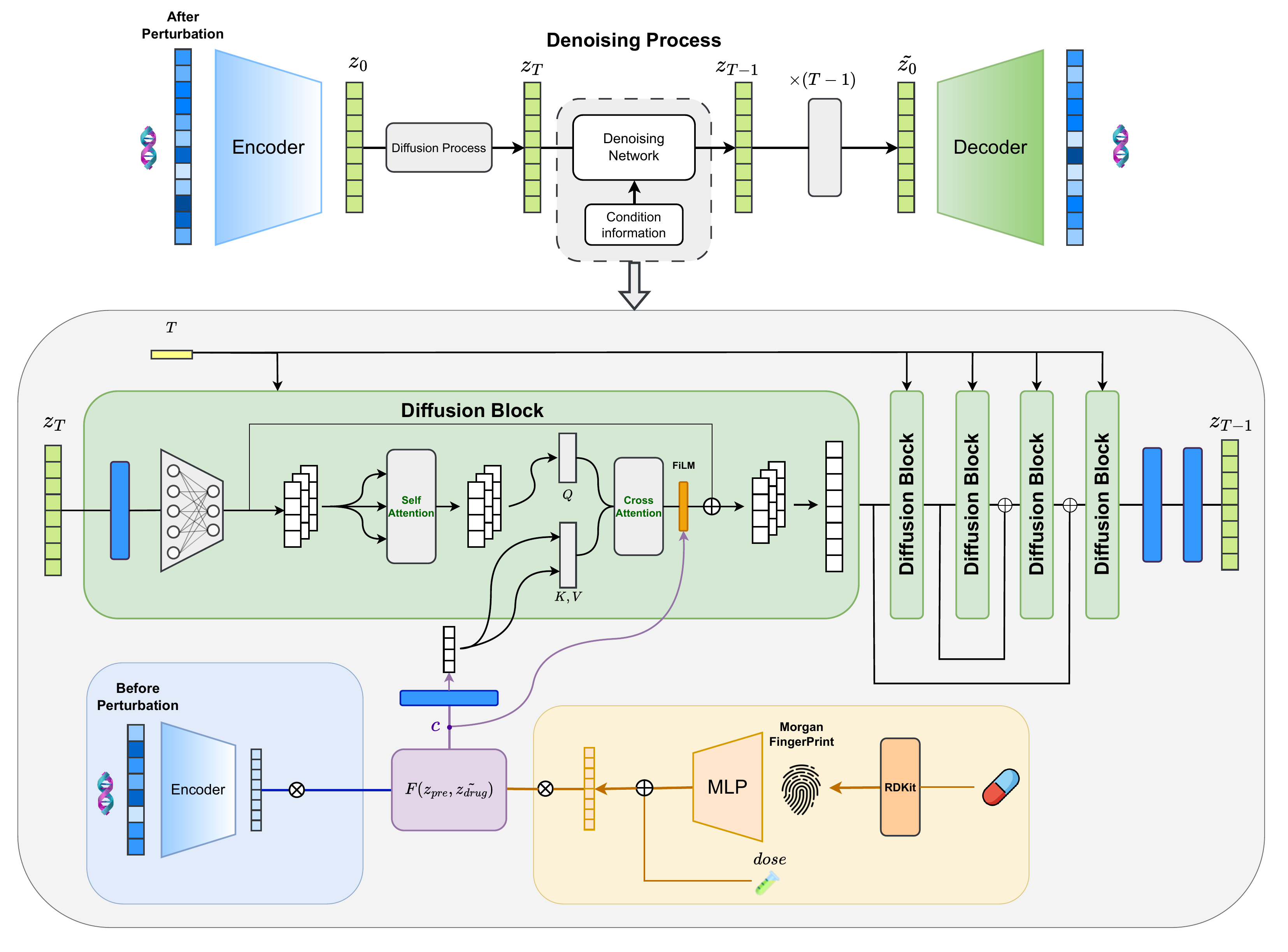}
  \caption{Overview of scTREAT. Top: Expression is encoded in a shared autoencoder latent space; the denoising path summarizes neural noise prediction and reverse-sampling updates to $\hat z_0$, followed by decoding to $\hat x$. Bottom: The matched-control representation $z_{\text{pre}}{=}E(x_{\text{pre}})$ and dose-fused drug vector $\tilde z_{\text{drug}}$ from Morgan fingerprints + MLP \citep{Rogers2010ECFP,LandrumRDKit} form a condition token $c$ injected through GD-Attn (\cref{sec.3.2}).}
  \label{fig1}
\end{figure*}

\paragraph{Approach and contributions.}
We present scTREAT (\cref{fig1}) and make two contributions:
\begin{itemize}
  \item \textbf{The first diffusion framework to factorize state and drug guidance for single-cell drug-response prediction.} A shared latent space and GD-Attn combine baseline expression with a dose-conditioned molecular representation. Four-state training supplies conditional predictions for factorized inference, with dose also setting drug-guidance strength.
  \item \textbf{Evidence from ablation and generalization.} We examine condition injection, molecular representation, and guidance through ablations, alongside UC/UD prediction on unobserved combinations and held-out drugs.
\end{itemize}

\section{Related Work}
\label{app:related}

\textbf{Deep learning for single-cell perturbation prediction.}
Latent-space generative approaches perform counterfactual prediction by mapping expression into a shared latent space: scGen uses latent vector arithmetic across cell types~\citep{Lotfollahi2019scGen}; CPA represents cellular background and dose as additive components~\citep{Lotfollahi2021.04.14.439903}; chemCPA further incorporates molecular structure encoders and transfer from bulk HTS to handle previously unseen compounds at single-cell resolution~\citep{hetzel2022predicting,Subramanian2017L1000}.
Structure-conditional generators combine chemical structure (e.g., SMILES/fingerprints \citet{Weininger1988SMILES,Schwartz2013SMIfp}), dose, and control expression to predict responses for untested compounds~\citep{Qi2024PRnet}.
Large-scale Transformer pretraining learns transferable gene-expression representations that can serve as initialization or encoders for downstream prediction and generation~\citep{Cui2024scGPT}.
Another line of latent representation learning is exemplified by BioLORD \citep{Piran2024Biolord}, which trains cross-modal encoders linking molecular structure and transcriptomic responses, enabling generalization to previously unseen compounds; in contrast, SAMS-VAE \citep{Bereket2024SAMS_VAE} models perturbation effects only for the perturbations observed during training and does not support unseen-drug prediction.
As an alternative to latent generative modeling, optimal-transport (OT) methods formulate control to perturbation alignment from unpaired observations to capture population-level shifts~\citep{Bunne2023NeuralOT,Dong2023CINEMAOT}.
A hybrid line combines large Transformers with OT, exemplified by STATE~\citep{Adduri2025.06.26.661135}, though its generative formulation does not support prediction for previously unseen compounds.

\textbf{Diffusion and transferable techniques from computer vision.}
In computer vision, diffusion provides an effective, scalable recipe for conditional generation: latent-space denoising with cross-attention~\citep{Rombach2022LDM}; controllable sampling via classifier or classifier-free guidance and compositional guidance~\citep{HoSalimans2022CFG,Liu2022Composable}; Transformer backbones with feature-wise modulation (AdaLN/FiLM)~\citep{Vaswani2017Attention,PeeblesXie2023DiT,Perez2018FiLM}; and control adapters for selective layerwise conditioning~\citep{Zhang2023ControlNet,Mou2023T2IAdapter}. More broadly, diffusion injects noise via a forward SDE and denoises via the reverse SDE or the probability-flow ODE~\citep{Song2020SDE}, with accelerated non-Markovian sampling such as DDIM~\citep{Song2020DDIM}.
For single-cell expression, scDiffusion~\citep{Luo2024scDiffusion} combines latent-space denoising with a pretrained encoder and uses classifier guidance and gradient-based conditional interpolation at inference.
Related diffusion approaches model perturbations through conditional generation or latent-space manipulation: scDiff formulates multiple single-cell tasks under a conditional diffusion model; scVAEDer combines a variational autoencoder with latent diffusion; Squidiff uses semantic latent representations for developmental, genetic, and drug perturbations; and dbDiffusion conditions latent generation on perturbation embeddings~\citep{Tang2023scDiff,Sadria2025scVAEDer,He2026Squidiff,Shang2025dbDiffusion}. Together, these studies establish diffusion as a viable tool for single-cell perturbation modeling. scTREAT builds on this foundation by combining shared latent conditioning with four-state training and an explicit rule for separate state and drug guidance. Its methodological focus is this combination and its evaluation through component ablations.

\textbf{Data resource.}
Large-scale resources such as LINCS/L1000 and Tahoe-100M support response modeling across compounds, doses, and cellular contexts. We use Tahoe-100M~\citep{Zhang2025Tahoe100M}, whose public release contains more than 100 million cells across 50 cancer cell lines and 379 drugs.

\section{Method}
\label{sec.3}
scTREAT models post-treatment expression conditioned on a baseline profile, molecular structure, and dose. A shared expression space connects baseline and response profiles; GD-Attn injects state and drug conditions into the denoiser; and four-state training supplies the predictions used for factorized guidance (\cref{fig1}).

\subsection{Latent-Space Diffusion Backbone}\label{sec.3.1}
We model the post-perturbation expression distribution in an observational manner as the conditional mapping
\begin{equation} \label{eq.1}
    p(x_{\text{post}} \mid x_{\text{pre}}, m, d),
\end{equation}
where $x_{\text{pre}}, x_{\text{post}} \in \mathbb{R}^{G}$ denote the pre- and post-perturbation expression over $G$ genes, $m$ denotes the drug's molecular structure (SMILES), and $d$ denotes dose.

A SCimilarity autoencoder \citep{Heimberg2023.07.18.549537}, fine-tuned following scDiffusion \citep{Luo2024scDiffusion}, provides an encoder $E$ and decoder $D$ that define a latent space of dimension $D_z$ in which the response dynamics are learned. Once fine-tuned, $E$ and $D$ are frozen. Post-treatment condition-level profiles are mapped to $z_0 = E(x_{\text{post}})$ for training, and the decoder reconstructs predictions via $\hat{x} = D(\hat{z}_0)$ after denoising.

Within this latent schema, a variance-preserving diffusion process injects Gaussian noise along a schedule $\{\beta_t\}_{t=1}^T$ with $\alpha_t = 1 - \beta_t$ and cumulative product $\bar{\alpha}_t = \prod_{s \le t} \alpha_s$. The forward transition is
\begin{equation}
    q(z_t \mid z_0) = \mathcal{N}\!\big(\sqrt{\bar{\alpha}_t}\, z_0,\, (1 - \bar{\alpha}_t)\, \mathbf{I}\big).
\label{eq.2}
\end{equation}
The denoiser $B_\theta$ receives the noisy latent $z_t$, a sinusoidal time embedding $\gamma(t) \in \mathbb{R}^{d_t}$, and condition $c$, and predicts the Gaussian noise $\hat{\epsilon}_t = B_\theta\big(z_t, \gamma(t),c\big)$. GD-Attn injects $c$ within the denoising blocks; the effect of this conditioning design is examined by the reported ablations.

\subsection{Condition Representation and GD-Attn Injection}\label{sec.3.2}
We encode the pre-perturbation state with the shared expression encoder and build a dose-aware, structure-informed drug vector before fusing both through GD-Attn.

\textbf{State encoding.} We reuse the same fine-tuned expression encoder $E$ of the backbone on the baseline state conditioning branch.
\begin{equation}
    z_{\text{pre}} = E^{\text{cond}}(x_{\text{pre}})\in\mathbb{R}^{D_{\text{pre}}},
    \quad D_{\text{pre}}=D_z \label{eq.4}
\end{equation}
$z_{\text{pre}}$ is the conditioning-side latent for the baseline state; $D_{\text{pre}}$ matches the backbone latent dimension $D_z$; $E^{\text{cond}} \equiv E$ indicates full parameter sharing.

The shared encoder provides a deterministic embedding in the same coordinates as the training target:
\begin{equation}
z_{\text{pre}}^{\text{cond}} = E(x_{\text{pre}}) = z_{\text{pre}}. \label{eq.5}
\end{equation}
Parameter sharing places baseline conditions and post-treatment profiles in a common expression space.

\textbf{Drug encoding.} Molecular structure informs the response representation, while dose modulates the drug condition. SMILES are converted to 1024-bit Morgan fingerprints $\phi\in\{0,1\}^{1024}$ and embedded via an MLP to obtain the structure-aware vector
\begin{equation}
    z_{\text{drug}}=\operatorname{MLP}(\phi)\in\mathbb{R}^{D_{\text{drug}}}\label{eq.6}
\end{equation}
where $D_{\text{drug}}$ is the dimensionality of the learned drug embedding. Dose is fused through a FiLM module with a continuous encoding $e_{\text{dose}}\in\mathbb{R}^{D_{\text{dose}}}$, where $D_{\text{dose}}$ denotes the dose-embedding dimension.
\begin{equation}
    \tilde z_{\text{drug}}=\operatorname{FiLM}\big(z_{\text{drug}},\,e_{\text{dose}}\big)\label{eq.7}
\end{equation}
Here $\operatorname{FiLM}$ applies dose-dependent feature-wise scaling and shifting to $z_{\text{drug}}$. Molecular information and dose jointly determine $\tilde z_{\text{drug}}$; this design allows dose to alter the representation without assuming an exact separation of response direction and magnitude.

\textbf{Condition injection.} Guided Decomposable Attention (GD-Attn) performs the non-concatenative injection. The composite condition is
\begin{equation}
    c = F\!\big(z_{\text{pre}},\, \tilde z_{\text{drug}}\big)\in\mathbb{R}^{D_c} \label{eq.8}
\end{equation}
$F$ is the fusion function and $D_c$ is the dimensionality of the condition.

GD-Attn refines latent tokens through self-attention, injects the fused condition through cross-attention to a single condition token, and applies FiLM modulation and a residual feed-forward update with LayerNorm. This pathway supplies cellular background and treatment information throughout denoising. Channel masking provides the condition subsets used for separate state/drug guidance.

\subsection{Four-State Training}\label{sec.3.3}
\label{sec.3.3.1}

Four-state training supplies the conditional predictions needed for separate state and drug guidance at inference.
We apply independent dropout to $\{z_{\text{pre}},\,\tilde z_{\text{drug}}\}$ in the training process to obtain the four possible combinations.
\begin{equation}
    c' \in \{\text{both},\ \text{no-pre},\ \text{no-drug},\ \varnothing\} \label{eq.9}
\end{equation}

Unified training objective is shown below.
\paragraph{Injection-side alignment and contrast.}
For each sample $i$, define the normalized empirical response direction
\begin{equation}
    \hat v^{(i)}=\frac{x_{\text{post}}^{(i)}-x_{\text{pre}}^{(i)}}{\max\!\big(\|x_{\text{post}}^{(i)}-x_{\text{pre}}^{(i)}\|_2,\varepsilon_v\big)}.
\end{equation}
We project the dose-fused embedding $\tilde z_{\text{drug}}$ with a learnable matrix $W\in\mathbb{R}^{G\times D_{\text{drug}}}$ and normalize the projected direction as
\begin{equation*}
    \hat u^{(i)}=\frac{W\,\tilde z_{\text{drug}}^{(i)}}{\max\!\big(\|W\,\tilde z_{\text{drug}}^{(i)}\|_2,\varepsilon_u\big)}.
\end{equation*}
Small positive constants $\varepsilon_v$ and $\varepsilon_u$ safeguard normalization. We align $\hat u^{(i)}$ to $\hat v^{(i)}$ while maintaining contrast with other response embeddings. The alignment loss is
\begin{equation}
\mathcal{L}_{\mathrm{align}}
=\big\|\hat v^{(i)}-\hat u^{(i)}\big\|_2^2.
\label{eq.10}
\end{equation}
To encourage discriminative drug--response representations, we use an InfoNCE term \citep{Oord2018CPC,Chen2020SimCLR}. With $s(u,v)=\langle u,v\rangle/\tau$ and negative response embeddings $\mathcal{N}_i$, the objective is
\begin{equation}
\mathcal{L}_{\mathrm{info}}
= \mathbb{E}_{i\in\mathcal{B}}\!\left[-\log
\frac{\exp\!\big(s(\hat u^{(i)},\hat v^{(i)})\big)}
{\exp\!\big(s(\hat u^{(i)},\hat v^{(i)})\big)+\sum_{v^-\in\mathcal{N}_i}\exp\!\big(s(\hat u^{(i)},v^-)\big)}\right].
\label{eq.11}
\end{equation}
Here $\tau$ is the temperature and $\mathcal{B}$ is the minibatch. In the in-batch form, $\mathcal{N}_i$ contains the other samples' response embeddings. The implementation also provides a module that draws negatives from a queue of detached response embeddings from previous batches, falling back to in-batch negatives while the queue is empty. This objective contrasts matched drug--response pairs with negative response samples.

The injection regularizer combines them as
\begin{equation}
\mathcal{L}_{\mathrm{inject}}
= \lambda_{\mathrm{align}}\,\mathcal{L}_{\mathrm{align}}
+ \lambda_{\mathrm{info}}\,\mathcal{L}_{\mathrm{info}}
\label{eq.12}
\end{equation}

\paragraph{Denoising loss under four-state training.}
With independent channel dropout $c'$,
the diffusion denoiser is trained by the MSE on noise:
\begin{equation}\label{eq.13}
\mathcal{L}_{\mathrm{denoise}}
= \mathbb{E}_{z_0,\,t,\,\epsilon,\,c'}\!\left[\big\|\epsilon - \epsilon_\theta(z_t, t, c')\big\|_2^2\right]
\end{equation}
where $z_t\!\sim\!q(z_t\mid z_0)$ follows the VP forward process and $\epsilon_\theta$ predicts the noise.

\paragraph{Total objective.}
Our overall objective weights the denoising criterion with the injection regularizer:
\begin{equation}\label{eq.14}
\mathcal{L}_{\mathrm{total}}
= \mathcal{L}_{\mathrm{denoise}}
+ \mathcal{L}_{\mathrm{inject}}
\end{equation}
with $\lambda_{\mathrm{align}},\lambda_{\mathrm{info}}\!\ge\!0$ as hyperparameters.

\subsection{Factorized Guidance}\label{sec.3.3.2}
At inference, we combine three conditional predictions. The coefficients $s_p$ and $s_d$ weight the state and drug guidance terms, respectively:
\begin{equation}
\label{eq.15}
\begin{split}
    \hat{\epsilon} = \; (1+s_p+s_d)\,\epsilon(&\text{both}) - s_p\,\epsilon(\text{no-pre})- s_d\,\epsilon(\text{no-drug}) \\
      & \quad s_p,s_d \ge 0
\end{split}
\end{equation}
$\epsilon(\cdot)$ denotes the network output under each condition. The unconditional state participates in training but is not evaluated in this inference rule. \Cref{eq.15} starts from $\epsilon(\text{both})$ and adds the state contrast $\epsilon(\text{both})-\epsilon(\text{no-pre})$ weighted by $s_p$ and the drug contrast $\epsilon(\text{both})-\epsilon(\text{no-drug})$ weighted by $s_d$. Thus, three predictions from the same denoiser suffice to construct the two guidance terms.

\begin{figure}[!t]
  \centering
  \includegraphics[width=0.68\linewidth,trim=140bp 0 0 0,clip]{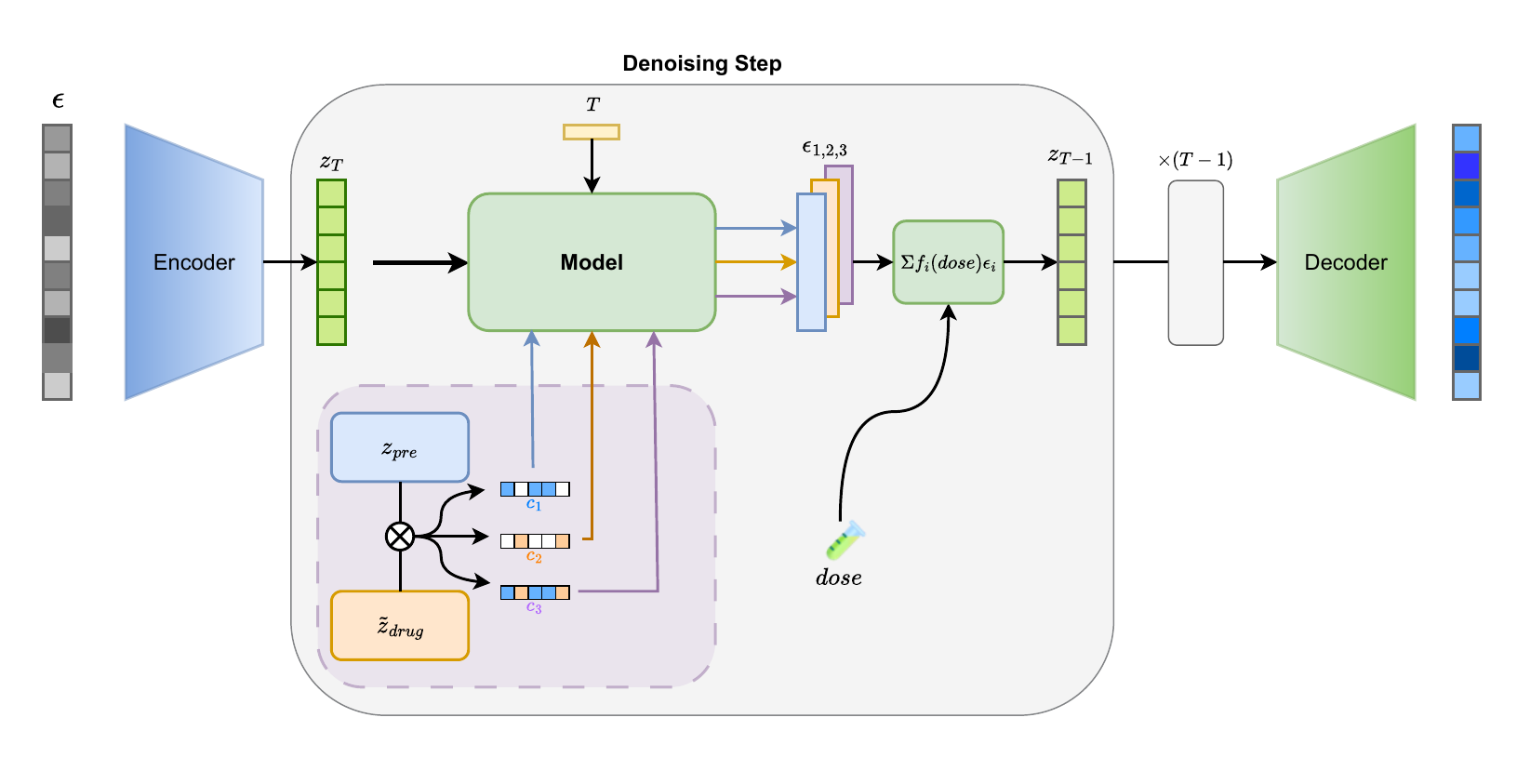}
  \caption{Conditional denoising from $z_T$. Three noise predictions from a four-state-trained denoiser are combined with separate state/drug guidance for each reverse-sampler update to $z_{t-1}$. Decoding the final $\hat z_0$ gives the predicted expression.}
  \label{fig2}
\end{figure}

We set drug-guidance strength continuously from dose:
\begin{equation}
    s_d(\text{dose}) = s_0 \cdot \sigma\!\big(\alpha \log(1+d) + \gamma\big),
    \qquad
    \sigma(z)=\tfrac{1}{1+e^{-z}} \label{eq.16}
\end{equation}
Here $d$ is dosage and $s_0,\alpha,\gamma$ are hyperparameters. The log-dose sigmoid \citep{Sebaugh2011EC50IC50} is continuous at zero and prevents unbounded amplification at high doses; the inference flow is shown in \cref{fig2}, with additional details in \cref{app:G}.

\section{Experiments}
We evaluate response prediction on unseen cell line--drug combinations (UC) and unseen drugs (UD), then examine condition injection, molecular representation, and guidance through component ablations.

\subsection{Benchmark and Training Protocol}
\paragraph{Data and splits.}\label{sec.4.1.1}
Tahoe-100M \citep{Zhang2025Tahoe100M} contains 50 cancer cell lines and 379 drugs in its public release. We average single-cell expression within each $(\text{cell line},\text{drug},\text{dose},\text{plate})$ group and match every non-DMSO condition to its same-plate DMSO control. UC holds out complete cell line--drug pairs and UD holds out complete drugs, grouping all doses and plates together (\cref{APP.E}).

\paragraph{Metrics.}\label{sec.4.1.2}
Metrics cover logFC and $\Delta$ correlations, median explained variance, and response-gene overlap. DEG denotes the Top-200 genes by observed absolute logFC unless specified otherwise. In a global paired pseudobulk analysis, \textbf{84\% (UC)} and \textbf{79\% (UD)} of the Top-200 genes satisfy $\mathrm{FDR}\le 0.05$ after one-sample t-tests on per-pair logFC and Benjamini--Hochberg correction across all genes (\cref{APP.C}).

\paragraph{Baselines.}\label{sec.4.1.3}\label{app:baselines}
Our model and other models are trained on the same set and evaluated on UC and UD test sets. All models use the same split, and each is trained to convergence. Baselines include Linear Regression and Multi-layer Perceptron (MLP) with drug encodings implemented analogously to PRnet \citep{Qi2024PRnet} adapter. The $\overline{\text{Perturb}}$ \citep{Adduri2025.06.26.661135} baseline averages observed perturbation offsets for the same perturbation in training data and adds mean offset to control expression to form prediction. The $\overline{\text{Context}}$ \citep{Adduri2025.06.26.661135} baseline averages perturbed expression across samples within the same cell line/context and uses this mean as prediction directly. We further compare six deep-learning baselines: chemCPA \citep{hetzel2022predicting}, an additive latent-space model with molecular structure encoding and dose scaling, supporting tests on both UC and UD; PRnet \citep{Qi2024PRnet}, an encoder--decoder model conditioned on drug perturbations via adapter that generates fingerprint-based embeddings, supporting both tasks above; BioLORD \citep{Piran2024Biolord} trains cross-modal encoders linking molecular structure and transcriptomic profiles and therefore supports both UC and UD evaluation; SAMS-VAE \citep{Bereket2024SAMS_VAE} models perturbation labels as discrete conditions without molecular-structure encoding and thus supports UC evaluation but cannot generalize to unseen drugs and is not evaluated on UD; STATE \citep{Adduri2025.06.26.661135}, a model that takes unperturbed transcriptome together with the perturbation label, evaluated only on UC and not supporting UD prediction; and scGPT \citep{Cui2024scGPT}, a Transformer-based single-cell foundation model used as perturbation-prediction baseline, also evaluated only on UC without supporting UD prediction.
All models, including baseline methods, are trained and evaluated on the same pseudobulked condition-level pairs for fairness.

\paragraph{Training.}
Only training-set profiles are used to fine-tune the SCimilarity autoencoder and train the diffusion model. Diffusion is optimized with AdamW \citep{Loshchilov2017_AdamW}; the learning rate warms up for 3{,}000 steps and then decays linearly. All models use the same split and condition-level pairs and are trained to convergence (\cref{APP.D,app:baselines}).

\subsection{Prediction on Unseen Combinations and Drugs}
\label{sec.4.2.1}
\begin{table}[!t]
\centering
\small
\setlength{\tabcolsep}{3pt}
\caption{UC results. Higher is better; bold and underline mark the best and second-best distinct values, with ties sharing formatting. Correlations and median explained variance ($EV$) are reported for all genes and the Top-200 genes by observed absolute logFC (DEG). P and S denote Pearson and Spearman correlations. Overlap@200 and Overlap@1000 denote the corresponding DEG-accuracy metrics.}
\label{tab:setA_combo}
\textbf{All genes}\par\smallskip
\begin{tabular*}{\textwidth}{@{\extracolsep{\fill}}lccccc@{}}
\toprule
Model & logFC-P & logFC-S & $\Delta$-P & $\Delta$-S & $EV_{\mathrm{median}}$ \\
\midrule
scTREAT & \cellcolor{lm_purple}\textbf{0.81} & \cellcolor{lm_purple}\textbf{0.69} & \cellcolor{lm_purple}\underline{0.55} & \cellcolor{lm_purple}\underline{0.51} & \cellcolor{lm_purple}\textbf{0.73} \\
chemCPA & 0.70 & 0.46 & \textbf{0.59} & 0.44 & \underline{0.70} \\
PRnet & 0.61 & \underline{0.58} & 0.35 & 0.27 & 0.51 \\
STATE & \underline{0.77} & \textbf{0.69} & 0.50 & 0.50 & 0.69 \\
scGPT & 0.46 & 0.39 & 0.39 & 0.35 & 0.41 \\
Linear & 0.50 & 0.51 & 0.48 & 0.41 & 0.20 \\
MLP & 0.55 & 0.57 & 0.44 & 0.36 & 0.40 \\
$\overline{\mathrm{Perturb}}$ & 0.46 & 0.38 & 0.21 & 0.25 & 0.48 \\
$\overline{\mathrm{Context}}$ & 0.48 & 0.32 & 0.25 & 0.24 & 0.39 \\
BioLORD & 0.50 & 0.34 & 0.46 & 0.37 & 0.63 \\
SAMS-VAE & 0.54  & 0.57  & 0.53  & \textbf{0.55}  & 0.53  \\
\bottomrule
\end{tabular*}
\par\medskip
\textbf{DEG and response-gene overlap}\par\smallskip
\begin{tabular*}{\textwidth}{@{\extracolsep{\fill}}lccccccc@{}}
\toprule
Model & logFC-P & logFC-S & $\Delta$-P & $\Delta$-S & $EV_{\mathrm{median}}$ & \shortstack{Overlap\\@200} & \shortstack{Overlap\\@1000} \\
\midrule
scTREAT & \cellcolor{lm_purple}\textbf{0.56} & \cellcolor{lm_purple}\textbf{0.49} & \cellcolor{lm_purple}\textbf{0.57} & \cellcolor{lm_purple}\textbf{0.59} & \cellcolor{lm_purple}\textbf{0.64} & \cellcolor{lm_purple}\textbf{0.24} & \cellcolor{lm_purple}\textbf{0.39} \\
chemCPA & 0.40 & 0.38 & 0.52 & 0.46 & \underline{0.59} & \underline{0.22} & 0.28 \\
PRnet & 0.31 & 0.28 & 0.36 & 0.28 & 0.37 & 0.11 & 0.21 \\
STATE & \underline{0.50} & \underline{0.48} & \underline{0.54} & \underline{0.52} & 0.53 & \underline{0.22} & \underline{0.32} \\
scGPT & 0.32 & 0.23 & 0.38 & 0.33 & 0.34 & 0.11 & 0.19 \\
Linear & 0.38 & 0.34 & 0.45 & 0.41 & 0.18 & 0.14 & 0.19 \\
MLP & 0.37 & 0.23 & 0.43 & 0.36 & 0.30 & 0.17 & 0.18 \\
$\overline{\mathrm{Perturb}}$ & 0.35 & 0.33 & 0.22 & 0.24 & 0.38 & 0.16 & 0.17 \\
$\overline{\mathrm{Context}}$ & 0.35 & 0.32 & 0.24 & 0.25 & 0.33 & 0.13 & 0.15 \\
BioLORD & 0.43 & 0.29 & 0.38 & 0.31 & 0.51 & 0.21 & 0.30 \\
SAMS-VAE & 0.44  & 0.45  & 0.46  & 0.49  & 0.44  & 0.20  & 0.29  \\
\bottomrule
\end{tabular*}
\end{table}

\begin{table}[!t]
\centering
\small
\setlength{\tabcolsep}{3pt}
\caption{UD results. Methods without unseen-drug support are omitted. Higher is better; bold and underline mark the best and second-best distinct values, with ties sharing formatting. Metric abbreviations and the DEG subset follow \cref{tab:setA_combo}.}
\label{tab:setA_drug}
\textbf{All genes}\par\smallskip
\begin{tabular*}{\textwidth}{@{\extracolsep{\fill}}lccccc@{}}
\toprule
Model & logFC-P & logFC-S & $\Delta$-P & $\Delta$-S & $EV_{\mathrm{median}}$ \\
\midrule
scTREAT & \cellcolor{lm_purple}\textbf{0.67} & \cellcolor{lm_purple}\textbf{0.58} & \cellcolor{lm_purple}\underline{0.50} & \cellcolor{lm_purple}\textbf{0.49} & \cellcolor{lm_purple}\textbf{0.66} \\
chemCPA & 0.42 & 0.44 & \textbf{0.53} & \underline{0.43} & 0.59 \\
PRnet & \underline{0.59} & 0.48 & 0.33 & 0.25 & 0.42 \\
Linear & 0.43 & \underline{0.49} & 0.46 & 0.40 & 0.18 \\
MLP & 0.40 & 0.36 & 0.43 & 0.34 & 0.32 \\
$\overline{\mathrm{Context}}$ & 0.31 & 0.32 & 0.42 & 0.39 & 0.40 \\
BioLORD & 0.42  & 0.30  & 0.45  & 0.37  & \underline{0.60}  \\
\bottomrule
\end{tabular*}
\par\medskip
\textbf{DEG and response-gene overlap}\par\smallskip
\begin{tabular*}{\textwidth}{@{\extracolsep{\fill}}lccccccc@{}}
\toprule
Model & logFC-P & logFC-S & $\Delta$-P & $\Delta$-S & $EV_{\mathrm{median}}$ & \shortstack{Overlap\\@200} & \shortstack{Overlap\\@1000} \\
\midrule
scTREAT & \cellcolor{lm_purple}\textbf{0.51} & \cellcolor{lm_purple}\textbf{0.49} & \cellcolor{lm_purple}\textbf{0.53} & \cellcolor{lm_purple}\textbf{0.52} & \cellcolor{lm_purple}\textbf{0.56} & \cellcolor{lm_purple}\textbf{0.22} & \cellcolor{lm_purple}\textbf{0.36} \\
chemCPA & \underline{0.38} & \underline{0.36} & \underline{0.48} & \underline{0.41} & 0.47 & \underline{0.20} & 0.28 \\
PRnet & 0.30 & 0.28 & 0.34 & 0.28 & 0.31 & 0.07 & 0.19 \\
Linear & 0.32 & 0.26 & 0.40 & 0.37 & 0.11 & 0.12 & 0.17 \\
MLP & 0.29 & 0.26 & 0.41 & 0.33 & 0.25 & 0.14 & 0.16 \\
$\overline{\mathrm{Context}}$ & 0.31 & 0.32 & 0.43 & 0.39 & 0.30 & 0.15 & 0.15 \\
BioLORD & 0.30  & 0.26  & 0.29  & 0.27  & \underline{0.51}  & 0.19  & \underline{0.29}  \\
\bottomrule
\end{tabular*}
\end{table}

The full model attains the best value on 10 of 12 UC metrics and 11 of 12 UD metrics (\cref{tab:setA_combo,tab:setA_drug}). In UC, representative gains over the second-best model are \textbf{+13.46\%} on DEG $\Delta$-Spearman and \textbf{+12.00\%} on DEG logFC-Pearson. In UD, DEG logFC-Spearman/Pearson reaches 0.49/0.51 versus 0.36/0.38 for chemCPA (\textbf{+36.11\%}/\textbf{+34.21\%}). Top-1000 DEG-Accuracy improves by \textbf{+21.88\%}/\textbf{+24.14\%} in UC/UD. Together with the ablations, these comparisons connect the conditioning and guidance design to its performance on the response-generation task.

The gains are not uniform: chemCPA leads all-gene $\Delta$-Pearson in UC and UD, while SAMS-VAE leads all-gene $\Delta$-Spearman in UC. The results therefore support broad improvements among the evaluated criteria while retaining distinctions between recovery of fold changes, perturbation shifts, and ranked response genes.

\subsection{Ablation Study}
\label{sec.4.3}

\begin{figure}[!t]
  \centering
  \includegraphics[width=0.99\linewidth]{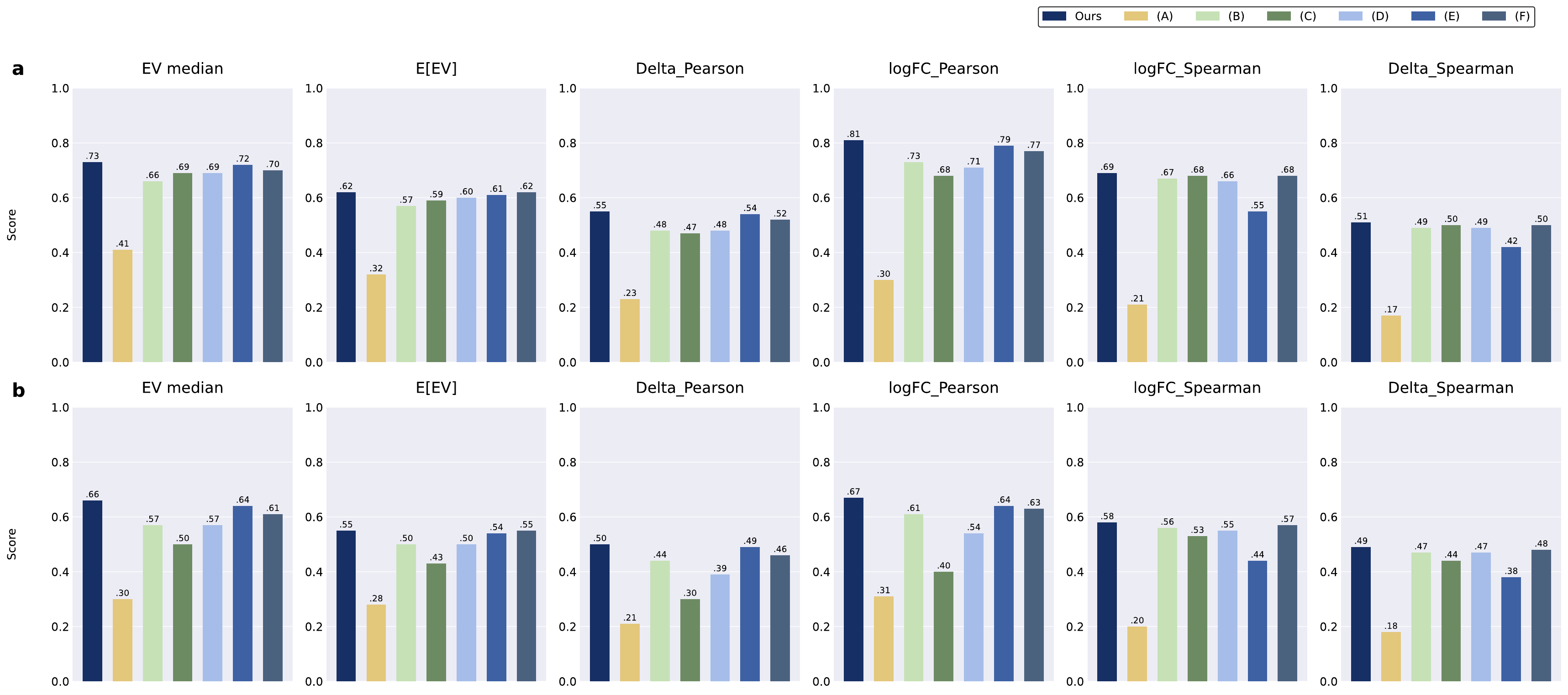}
  \caption{Ablations on UC (a) and UD (b). Conditioning, GD-Attn, and separate guidance support predictive performance; molecular structure supports UD generalization; dose-dependent guidance improves selected metrics. Variants: (A) no conditioning; (B) GD-Attn replaced by Concat+MLP; (C) molecular structure replaced by Drug-ID; (D) single-channel CFG; (E) constant drug-guidance strength; (F) concatenation of condition and latent tokens before attention. Full configurations are in \cref{app:F}.}
  \label{fig4}
\end{figure}

We examine how the conditioning and guidance design affects prediction (\cref{fig4}). Removing conditioning collapses performance; replacing GD-Attn weakens fit and $\Delta$ alignment; replacing molecular structure with Drug-ID most strongly harms UD; single-channel guidance lowers several metrics; and constant $s_d$ reduces predictive accuracy on selected criteria. Full configurations are in \cref{app:F}.

\subsection{Qualitative Prediction}
\begin{figure}[!t]
  \centering
  \includegraphics[width=0.94\linewidth]{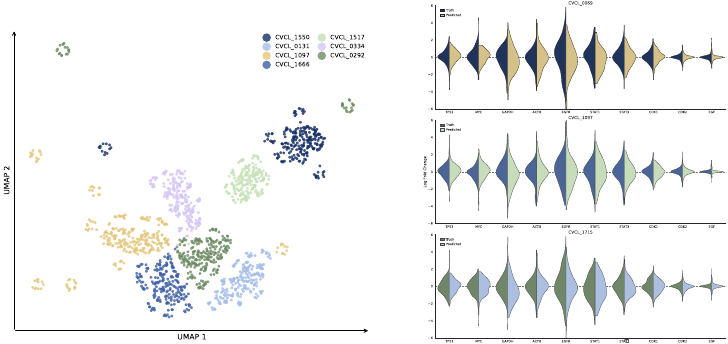}
  \caption{Qualitative response prediction. Predicted profiles separate by cell line in UMAP space (left), while predicted and observed gene-wise logFC distributions agree for representative cell lines (right).}
  \label{fig:prediction}
\end{figure}

The qualitative comparison complements the aggregate metrics by showing that predicted profiles retain cell-line structure and reproduce representative gene-wise logFC distributions.

\begin{figure}[!t]
  \centering
  \begin{minipage}[c]{0.49\linewidth}
    \centering
    \includegraphics[width=\linewidth]{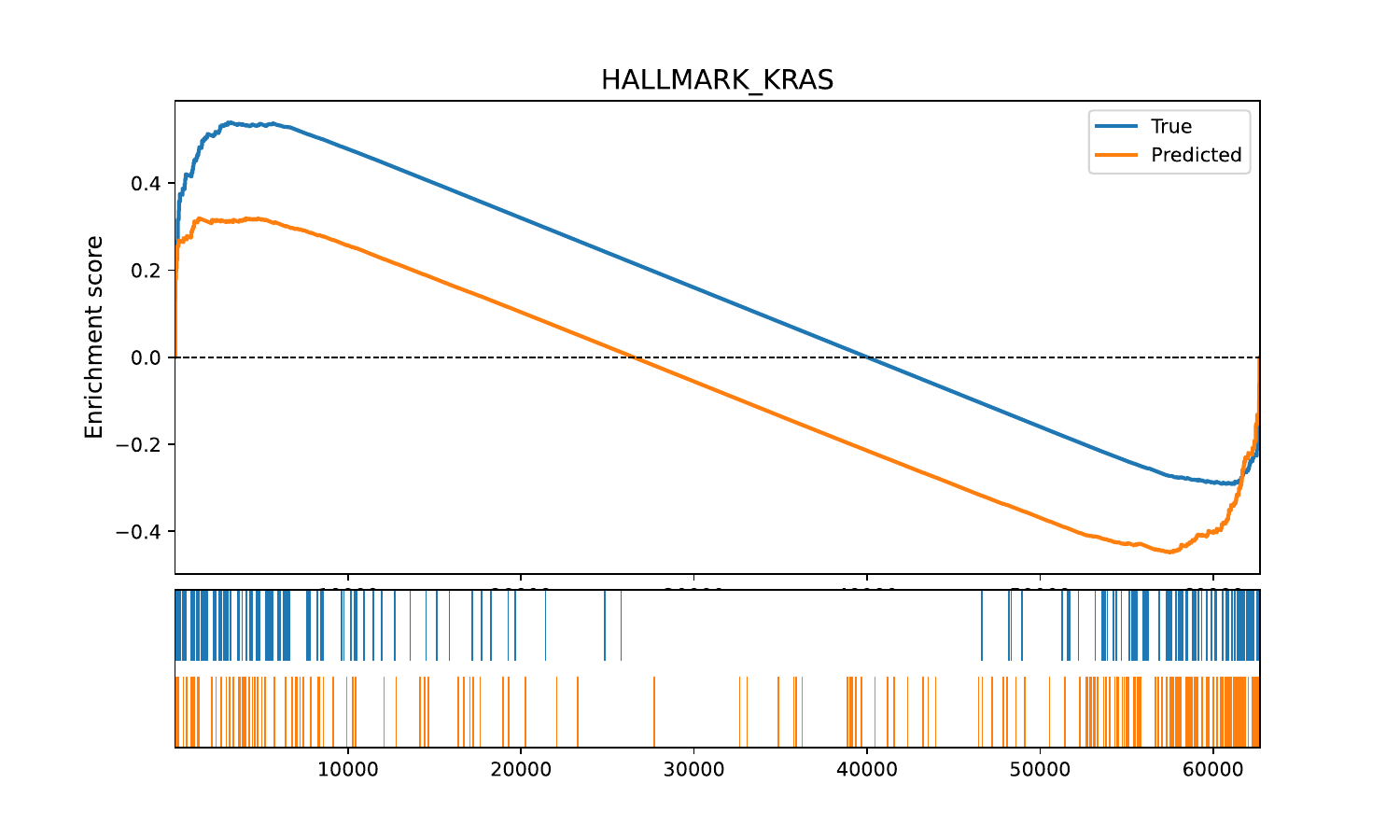}
  \end{minipage}\hfill
  \begin{minipage}[c]{0.43\linewidth}
    \centering
    \includegraphics[width=0.72\linewidth]{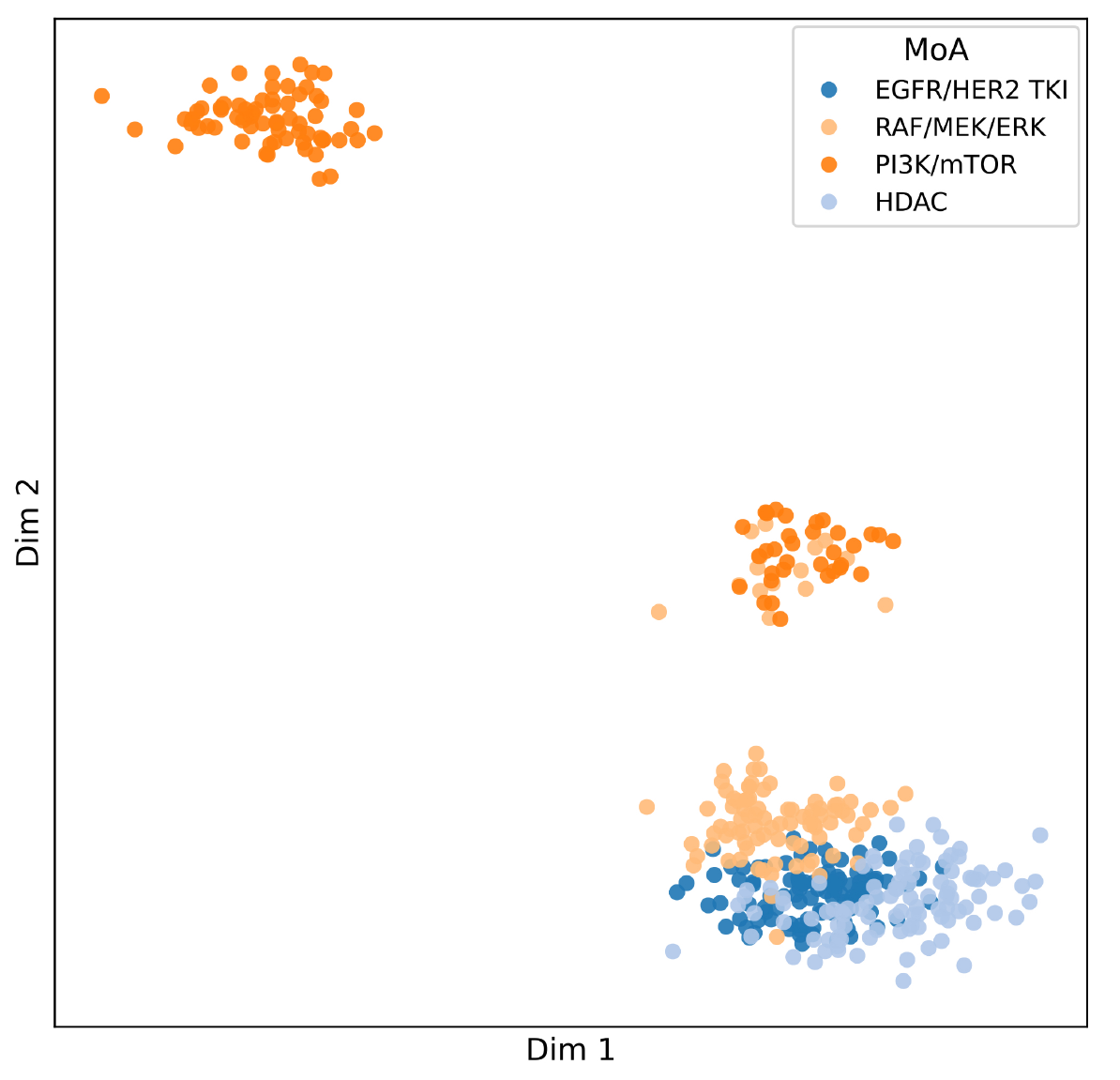}
  \end{minipage}
  \caption{Biological context. Left: KRAS GSEA curves share a broad shape but differ in their dominant signed deviations. Right: drug representations show MoA-associated grouping with overlap between some classes.}
  \label{fig:bio}
\end{figure}

\subsection{Biological Context}
\paragraph{Pathway analysis.}
The KRAS example in \cref{fig:bio} and the PI3K--AKT--mTOR example in \cref{fig5} show similar broad running-enrichment curve shapes: both predicted and observed profiles rise near the beginning of the ranked list and decline through its middle. These shared features provide a qualitative view of pathway-associated responses. The amplitudes and dominant signed deviations differ, however: the observed curves have larger positive peaks, whereas the predicted curves have larger negative troughs. Under the standard signed-ES interpretation, this limits agreement in enrichment direction, so the examples support partial correspondence rather than complete pathway-response recovery.

\paragraph{MoA-associated representation structure.}
The MoA-colored UMAP in \cref{fig:bio} shows pharmacologically associated organization, including a distinct region enriched for PI3K/mTOR compounds and localized groupings of other annotated classes. EGFR/HER2 and HDAC compounds also overlap. These patterns provide positive qualitative evidence of MoA-associated structure in the learned drug representation, without requiring complete separation of the annotated classes. Because molecular fingerprints are model inputs, this view alone does not establish mechanistic information beyond the supplied structural features.

\FloatBarrier
\section{Discussion and Conclusion}
scTREAT is the first diffusion framework to factorize state and drug guidance for single-cell drug-response prediction. It combines shared latent expression modeling, non-concatenative condition injection, and four-state training for this inference design. Ablations and UC/UD comparisons connect the design to predictive performance, with competing methods retaining advantages on selected $\Delta$ metrics. The results support latent diffusion for predicting transcriptional responses to unmeasured combinations and drugs, while the qualitative analyses provide complementary views of cellular and pharmacological structure.

\clearpage
\subsection*{AI Use Statement}
AI tools were used to polish writing, support literature retrieval and discovery, and assist with part of the research execution.

\subsection*{Reproducibility Statement}
The model formulation and training objective are specified in \cref{sec.3}, with baseline details in \cref{app:baselines}. The appendix describes data construction and split grouping (\cref{APP.E}), training (\cref{APP.D}), evaluation metrics (\cref{APP.C}), and ablation configurations (\cref{app:F}).

An anonymized ZIP archive containing the core code has been submitted as supplementary material.

\bibliographystyle{iclr2027_conference}
\bibliography{references}
\clearpage
\appendix
\section{Additional Context}
\paragraph{Experimental motivation.}
While enabling parallel observation of drug effects at some scale, traditional high-throughput screening (HTS) workflows \citep{Mayr2009_NovelTrendsHTS,Macarron2011HTSImpact,Iversen2012_HTSValidation_AGM} are extremely costly in terms of reagents, personnel time, and turnaround. Their throughput is also limited, making it infeasible to cover all potential cell line $\times$ drug $\times$ dose combinations; even within large public perturbation atlases, measured conditions constitute only a tiny fraction of the combinatorial space. Perturbation maps can additionally support the exploration of precision therapeutic strategies.

\section{Additional Inference Details}\label{app:G}
The coefficients $s_p$ and $s_d$ separately weight the state and drug contrast terms in \cref{eq.15}. Both terms share the fully conditioned prediction, so this interface does not require biological independence of the two conditions. Dose enters both the FiLM representation and the bounded log-dose sigmoid in \cref{eq.16}; the reported ablation measures predictive performance, without establishing monotonic gene-level responses.

The sampler starts from a latent state $z_T$. At each step, three neural noise predictions are combined by \cref{eq.15}, and the reverse sampler updates the latent. The expression encoder is used to obtain the baseline condition, and the decoder maps the final $\hat z_0$ back to expression.

\section{Training and Data Construction}

\subsection{Training Setup}
\label{APP.D}
Following scDiffusion~\citep{Luo2024scDiffusion}, we adapt the SCimilarity autoencoder \citep{Heimberg2023.07.18.549537} to the expression dimension and fine-tune it on training-set profiles with a reconstruction objective. We then freeze the encoder and decoder for latent diffusion training. Optimization uses AdamW \citep{Loshchilov2017_AdamW}; the diffusion learning rate warms up for 3{,}000 steps and then decays linearly. The held-out test sets are excluded from autoencoder fine-tuning, model training, and model selection.

\subsection{Dataset Construction and Splits}
\label{APP.E}
We use the Tahoe-100M \citep{Zhang2025Tahoe100M} public single-cell drug-perturbation atlas as our data source. This dataset is at hundred-million scale, spans 50 cancer cell lines, includes 379 drugs in its public release, and is compatible with the conditional mapping $(x_{\text{pre}},\, \text{drug},\, \text{dose}) \!\to\! x_{\text{post}}$ that we aim to learn. We take $(\text{cell\_line}, \text{drug}, \text{dose}, \text{plate})$ as conditioning keys, averaging single-cell expression within each key group to obtain the corresponding representative state. For each non-DMSO condition, we identify the matched DMSO control on the same plate to establish a one-to-one pair, yielding the pre-perturbation transcriptome $x_{\text{pre}}$ and post-perturbation transcriptome $x_{\text{post}}$.

UC holds out complete cell line--drug combinations: all doses and plates for a held-out combination remain outside training, while its cell line and drug each appear individually in training. UD holds out complete drugs across all cell lines, doses, and plates. These grouping rules prevent the held-out combination or drug from re-entering training through another dose or plate.

\section{Evaluation Metrics}
\label{APP.C}

We rank all genes by absolute log$_2$ fold change ($|\log_2\text{FC}|$) under the perturbation and take the top-$k$ genes as the high-impact set. For all DEG-based evaluations except (4), we fix $k=200$. We retain DEG as shorthand for this effect-size-selected response-gene subset; selection does not require passing an FDR threshold. We evaluate the following metrics, each computed on both the full gene set and the DEG subset:
(1) Pearson and Spearman correlations between the ground-truth and predicted log fold-change vectors ($\log_2\text{FC}$);
(2) Pearson and Spearman correlations between the ground-truth and predicted perturbation-shift vectors ($\Delta := x_{\text{post}} - x_{\text{pre}}$);
(3) median explained variance of the predictions with respect to the ground truth; and
(4) DE-Overlap-Accuracy, for which we evaluate several values of $k$.

As an additional statistical check, we conducted a global paired pseudobulk analysis separately for UC and UD. For each gene, we performed a one-sample t-test on its per-pair logFC values against a zero mean, followed by Benjamini--Hochberg correction across all genes. Among the Top-200 genes ranked by absolute logFC in this analysis, \textbf{84\% (UC)} and \textbf{79\% (UD)} satisfy $\mathrm{FDR}\le 0.05$. Thus, a large majority of the selected genes also pass the significance threshold. This global check complements the effect-size-based evaluation; it does not redefine the evaluated subset or test performance differences between models.

\section{Additional Pathway Example}
\label{app:additional_validation}

\begin{figure}[!htbp]
  \centering
  \includegraphics[width=0.9\linewidth]{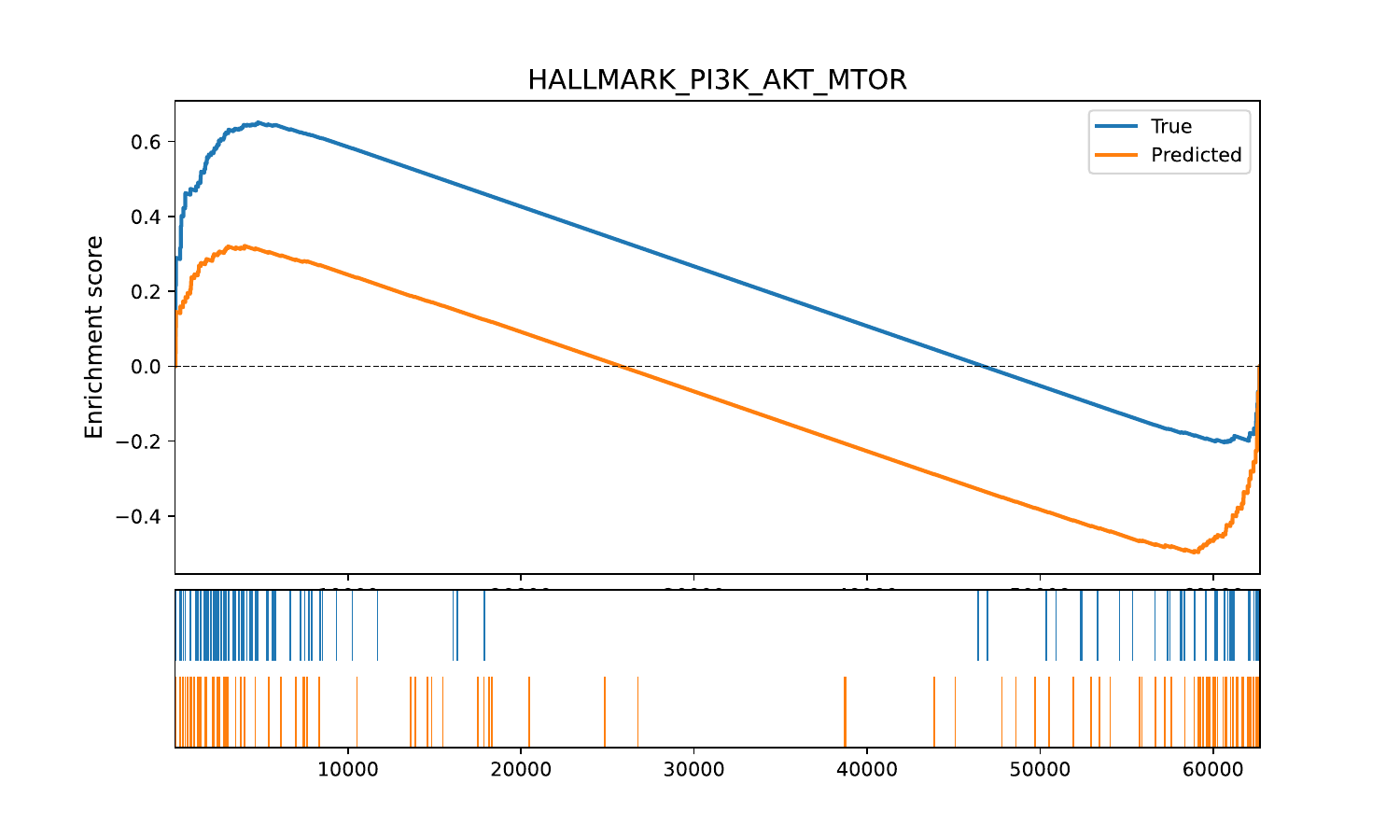}
  \caption{Exploratory GSEA comparison for HALLMARK\_PI3K\_AKT\_MTOR\_SIGNALING. Predicted and observed curves share broad shape features but differ in amplitude and dominant signed deviation.}
  \label{fig5}
\end{figure}

\Cref{fig5} compares the running enrichment scores for the PI3K--AKT--mTOR pathway computed from observed and predicted responses. Both curves rise near the beginning of the ranked list before declining. The predicted curve has a smaller positive peak and a deeper negative trough than the observed curve. Tick marks below the curves show where pathway genes occur in each ranked list. This example complements the KRAS analysis in \cref{fig:bio}, illustrating similarities in curve shape as well as differences in enrichment magnitude and pathway-gene ranking.

\FloatBarrier

\section{Complete Ablation Configuration}
\label{app:F}

We examine five design factors through six ablation variants:
(A) remove all conditioning (No-Cond);
(B) replace GD-Attn (\cref{sec.3.2}) with Concat+MLP;
(C) replace the Morgan/SMILES-based drug encoder (\cref{eq.6}) with Drug-ID;
(D) replace decomposable dual-knob guidance $(s_p, s_d)$ (\cref{eq.15}) with single-channel CFG;
(E) replace the dose-to-strength mapping $s_d(\mathrm{dose})$ (\cref{eq.16}) with a constant; and
(F) replace the non-concatenative GD-Attn injection (\cref{sec.3.2}) with an input-level concatenation before attention, which concatenates the fused condition token with the latent tokens prior to the attention blocks.

Across UC/UD, these ablations examine conditioning, GD-Attn injection, structure-aware drug encoding, separate guidance, and the dose-to-strength mapping. GD-Attn improves several fit and $\Delta$-alignment metrics relative to the concatenation variants, while separate and dose-dependent guidance improve selected predictive criteria. These comparisons assess prediction quality rather than directly measuring training stability or dose monotonicity; structure-aware priors are especially important for UD generalization.

\section{Additional Discussion}
The methodological focus of scTREAT is the connection between condition representation and the inference interface. State and drug information enter a shared denoiser through GD-Attn. Independent channel dropout trains the model under four conditioning states, and three conditional predictions are then combined using separately specified state and drug coefficients. This formulation makes the guidance rule explicit while retaining a common response-generation backbone.

The ablations examine different parts of this design: availability of conditions, the injection pathway, molecular representation, and the guidance rule. The UC/UD comparisons provide the task-level evaluation of the resulting generator. The advantages of competing models on selected all-gene $\Delta$ metrics show that the benefits depend on which aspects of the response are measured.

\end{document}